\documentclass[final,5p,times,twocolumn]{elsarticle}

\usepackage{amssymb}
\usepackage{xcolor}
\usepackage{multirow}
\usepackage{float}
\usepackage{stfloats} 
\usepackage{hyperref}
\hypersetup{%
    colorlinks=true,        % false: boxed links; true: colored links
    linkcolor=blue,          % color of internal links (change box color with linkbordercolor)
    citecolor=blue,         % color of links to bibliography
    urlcolor=blue           % color of external links
    }

\pdfoutput=1

\journal{Physics Letters B}

\begin{document}

\begin{frontmatter}



\title{Thermal effects on tidal deformability in the last orbits of an inspiraling binary neutron star system}


\author{A. Kanakis-Pegios}
\ead{alkanaki@auth.gr}
\author{P.S. Koliogiannis\corref{cor1}}
\ead{pkoliogi@physics.auth.gr}
\author{Ch.C. Moustakidis}
\ead{moustaki@auth.gr}

\cortext[cor1]{P.S. Koliogiannis}

\affiliation{organization={Department of Theoretical Physics, Aristotle University of Thessaloniki},%Department and Organization
            %addressline={}, 
            city={Thessaloniki},
            postcode={54124}, 
            %state={},
            country={Greece}}

\begin{abstract}
The study of binary neutron stars mergers by the detection of the emitted gravitational waves is one of the most promised tools to study the properties of dense nuclear matter at high densities. It is worth claiming that strong evidence that the temperature of the stars during the last orbits before coalescing is  very low, $T\ll 1$ MeV (hereafter cold case or cold star; ${\rm k}_{B}=1$), does not exist. Nevertheless, theoretical studies suggest that the temperature concerning the inspiral phase, could reach even  a few MeV. The heating  process  of the interior of the neutron stars is as follows; tides transfer mechanical energy and angular momentum to the star at the expense of the orbit, where friction within the star converts the mechanical energy into heat. During the inspiral, these effects are potentially detectable. Different treatments have been used to estimate the transfer of the mechanical energy and the size of the tidal friction, leading to different conclusions about the importance of pre-merger tidal effects. The present work is dedicated to the study of the effect of temperature on the tidal deformability of neutron stars during the inspiral of a neutron star system just before the merger. We applied a class of hot equations of state, both isothermal and adiabatic, originated from various nuclear models. We found that even for low values of temperature ($T<1$ MeV), the effects on the basic ingredients of tidal deformability are not negligible. On the other hand, in the case of the adiabatic star, the thermal effects on tidal deformability remain imperceptible, up to the entropy per baryon value $S=0.2$ (in units of ${\rm k}_{B}$). According to the main finding, the effects of the temperature on tidal deformability, at least for low temperatures, are almost negligible. This is a surprising effect since both radius and Love number, the basic ingredients of the tidal deformability, are sensitive to temperature. The consequences of the above  result are discussed and analyzed.

\end{abstract}


\end{frontmatter}

%%\linenumbers

%\linenumbers
%%%%%%%%%%%%%%%%%%%%%%%%%%%%%%%%%%%%%%%
\section{Introduction}
%%%%%%%%%%%%%%%%%%%%%%%%%%%%%%%%%%%%%%
The last years very rich information has been gained by the detection of gravitational waves of binary black holes, binary neutron stars, and black hole-neutron star systems by the LIGO/Virgo collaboration~\cite{Abbott-gw170817,Abbott-gw190425}. In particular, the inspiral phase of binary neutron stars system, as well as the merger and
post-merger epoch, are considered as ones of the best extraterrestrial
laboratories to study unexplored properties of dense cold and hot nuclear matter. This study is based on the comparison of  theoretical predictions to the observational constraints concerning the tidal
deformability~\cite{Radice-2020,Baiotti-2019,Chatziioannou-2020,Lasky-2021,Dietrich-2021,Kanakis-2020,Kanakis-2021,Kolliogiannis-2021b}.

It is worth noticing that the suggestion that the temperature of the stars is zero during the last orbits before coalescing, is not subject to strong evidence. Contrariwise, there are some theoretical statements or conjectures claiming that the temperature of the star might even be a few MeV~\cite{Meszaros-92,Bildsten-92,Kochanek-92,Most-2022,Perego-2019}. Following the very recent study of Arras and Weinberg~\cite{Arras-2019}, tides transfer mechanical energy and angular momentum to the star at the expense of the orbit, and then friction within the star converts the mechanical energy into heat. During the inspiral, these effects are potentially detectable as a deviation of the orbital decay rate from the general relativistic point-mass result, or as an electromagnetic precursor, if heating ejects the outer layers of the star. Different treatments have been used to estimate the transfer of energy and the size of the tidal friction, leading to different conclusions about the importance of the pre-merger tidal effects~\cite{Arras-2019}.

Firstly, Meszaros and Rees~\cite{Meszaros-92} stated that during the final spiraling of the binary neutron star system, the two stars will distort each other tidally. These effects lead to an appreciable heat before the merger. It is stated that due to the effects of tidal heating, the neutron star will heat up to temperatures (both the core and crust) comparable to the ones of supernova explosions~\cite{Meszaros-92}. Similar results have been found in Refs.~\cite{Bildsten-92,Kochanek-92}, while a relevant discussion has been given  in the  recent studies~\cite{Most-2022,Perego-2019}. Other studies concerning the tidal heating due to (a) resonant excitation of g-mode, (b) inertial moment resonances, (c) bulk and shear viscosity, (d) and superfluid effects, have been presented in Refs.~\cite{Lai-94,Reisenegger-94,Ho-99,Lai-2006,Xu-2017,Yu-2017}. In particular, in the notable study of Ref.~\cite{Lai-94}, the key role of the viscosity on the tidal heating was stressed out. According to this study the tidal dissipation heats up the neutron stars at most in the range $10^{-4}-10^{-2}$ MeV before the merger. Recently, Arras and Weinberg~\cite{Arras-2019} studied systematically  the impact of Urca reactions driven by tidally induced fluid motion during binary neutron star inspiral~\cite{Arras-2019}. They come to the conclusion that even the direct Urca process is not efficient to heat the core of the neutron star in temperatures higher than $10^{-2}$ MeV. Moreover, the crust meltdown in inspiraling binary neutron stars has been discussed by Pan \textit{et al.}~\cite{Pan-2020}. They concluded that this process could induce a small but not negligible phase shift in the gravitational waveform, which may be observed with Advanced LIGO Plus or the featured Einstein Telescope and Cosmic Explorer~\cite{Pan-2020}. Furthermore, a comprehensive study on the heat blanketing envelopes of neutron stars has been given in Ref.~\cite{Yakovlev-2021}. In any case, in each of the above studies, the main conclusion is the fundamental role of the value of viscosity in the matter of neutron stars. This value plays a key role in the amount of heating of a neutron star in the inspiral before the merger process.

All the above studies concluded that tidal heating effects, during the inspiral, are present (due to different reasons) and also lead to the heating of the interior of neutron stars. The predictions cover a large interval, higher than three orders of magnitude, $T=0.01-10$ MeV. The tidal deformability is a quantity sensitive to the equation of state (EoS) of neutron stars. In particular, it depends both on the tidal Love number $k_2$ and the radius $R$. Since both quantities are determined within the applied EoS and for a fixed value of the mass, it will be of great interest to study the deviation of both quantities from the cold case, due to the existence of temperature. This is important since the tidal deformability, and consequently the corresponding radius, are constrained from the detection of the gravitational waves.

The motivation of the present work is to examine to what extent the temperature, due to various mechanisms, affects the values of the tidal deformability of a neutron star during the inspiral process, just before the merger. According to our knowledge, this possibility has never been examined. We concentrate on the case of isothermal matter in the interior of the neutron stars, employing some of the most used EoSs, for temperatures in the range $T=0.01-1$ MeV~\cite{Lattimer-91,Shen-2011,Banik-2014,Steiner-2013,Koliogiannis-2021}.  Moreover, for reasons of completeness and comparison, we extend the study to the more extreme but less reliable cases of even higher temperatures. Details for the construction of hot EoSs, both for low and high nuclear matter densities, are given also in Refs.~\cite{Perego-2019,Lattimer-2016,Prakash-1997,Steiner-2019,Randrup-2009,Constantinou-2014,Moustakidis-2009,Moustakidis-2009b,Burgio-2021,Typel-2018,Typel-2010,Bonasera-2014,Raduta-2020,Raithel-2019,Kaplan-2014,Chamel-2020,Chamel-2011}. 

The main reason we are addressing this issue is that, at least up to now, there are no theoretical predictions about the effect of temperature on tidal deformability. We believe that this effect must be studied in detail. The plausibility of the present study could be compared to observational data (current and future ones) along with the thermodynamic state of the binary system of neutron stars.

Obviously, the study of the isothermal equilibrium of the star is not the most realistic. Nevertheless, it is a first order approach that is more realistic than considering a cold star. Clearly, at the temperatures at which the present study takes place, the structure of the star's core is not particularly affected. On the contrary, the crust is much more sensitive to temperature. However, in the area of the mass size which is mainly detected by gravitational waves (1.2-1.6 $M_{\odot}$), the radius and consequently the tidal deformability, are particularly sensitive to the structure and size of the crust. Moreover, in order to enrich our study, we use a set of adiabatic (isentropic) EoSs for entropy $S<1$. In this case, the neutron star is considered to be in an adiabatic equilibrium.

The article  is organized  as follows: In Sec.~\ref{sec:eos} we present the warm EoSs, where in Sec.~\ref{sec:tidal}, the basic definitions related to the tidal deformability are provided. The results, as well as their discussion, are explained in detail in Sec.~\ref{sec3:Results}. Finally, Sec.~\ref{sec3:Conclus} contains the conclusions of the present study.

%%%%%%%%%%%%%%%%%%%%%%%%%%%%%%%%%%%%%%%%%%%%%%%%%%%%%%%%%%%%%%%%%%%%%%%%%%%%%%%%%%%%%%%%%%%%%%%%%%%%%%%%%%%%%%%%
\section{Warm neutron star matter } \label{sec:eos}
%%%%%%%%%%%%%%%%%%%%%%%%%%%%%%%%%%%%%%%%%%%%%%%%%%%%%%%%%%%%%%%%%%%%%%%%%%%%%%%%%%%%%%%%%%%%%%%%%%%%%%%%%%%%%%%%
The EoSs for the interior of neutron stars at finite temperature, as well as entropy per baryon, utilize the work of Koliogiannis and Moustakidis~\cite{Koliogiannis-2021}, where the data for the energy per particle for the symmetric nuclear matter and pure neutron matter, regarding the APR-1 EoS~\cite{Akmal-1998}, and the momentum-dependent interaction model are employed. In particular, the above parameterization is applied for the construction of three isothermal EoSs with $T=[0.01,0.1,1]~{\rm MeV}$ in a beta equilibrium state, and three isentropic EoSs with $S=[0.1,0.2,0.5]$ and proton fraction $Y_{p}=0.2$.
%%%%%%%%%%%%%%%%%%%%%%%%%%%%%%%%%%%%%%%%%%%%%%%%%%%%%%%%%%%%%%%%%%%%%%%%%%%%%%%%%%%%%%%%%%%%%%%%%%%%%%%%%%%%%%%%
\subsection{Core of neutron stars}
%%%%%%%%%%%%%%%%%%%%%%%%%%%%%%%%%%%%%%%%%%%%%%%%%%%%%%%%%%%%%%%%%%%%%%%%%%%%%%%%%%%%%%%%%%%%%%%%%%%%%%%%%%%%%%%%
Considering the absence of neutrinos in adiabatic EoSs (to simulate the rising of temperature due to merger), in both profiles, isothermal and adiabatic, the key relation for the chemical potentials $\mu_{i}, \; i=\{e,p,n\}$ is
\begin{equation}
    \mu_{e} = \mu_{n} - \mu_{p} = 4I(n,T)\left[F(n,T,I=1) - F(n,T,I=0)\right],
    \label{eq:chem_pot}
\end{equation}
where $F(n,T,I)$ is the free energy per particle, and $I=1-2Y_{p}$ is the asymmetry parameter, assuming that the EoS contains protons, neutrons, and electrons. It has to be noted that the calculation of the proton fraction in isothermal profile is possible via the beta equilibrium state, the Eq.~(\ref{eq:chem_pot}), and the density of electrons defined as
\begin{equation}
    n_{e} = \frac{2}{\left(2\pi\right)^{3}} \int\frac{d^{3}k}{1+\exp\left({\frac{\sqrt{\hbar^{2}k^{2}c^{2}+m_{e}^{2}c^{4}}-\mu_{e}}{T}}\right)}.
\end{equation}

However, in adiabatic profile, we consider that the proton fraction is almost constant. Finally, the energy density and pressure for the core of neutron stars, are constructed as
\begin{eqnarray}
    \mathcal{E}(n,T,I) &= \mathcal{E}_{b}(n,T,I) + \mathcal{E}_{e}(n,T,I), \label{eq:energy_density} \\
    P(n,T,I) &= P_{b}(n,T,I) + P_{e}(n,T,I).
    \label{eq:pressure}
\end{eqnarray}

Therefore, Eqs.~(\ref{eq:energy_density}) and~(\ref{eq:pressure}) constitute the ingredients for the EoSs in hot nuclear matter.
%%%%%%%%%%%%%%%%%%%%%%%%%%%%%%%%%%%%%%%%%%%%%%%%%%%%%%%%%%%%%%%%%%%%%%%%%%%%%%%%%%%%%%%%%%%%%%%%%%%%%%%%%%%%%%%%
\subsection{Crust of neutron stars}
%%%%%%%%%%%%%%%%%%%%%%%%%%%%%%%%%%%%%%%%%%%%%%%%%%%%%%%%%%%%%%%%%%%%%%%%%%%%%%%%%%%%%%%%%%%%%%%%%%%%%%%%%%%%%%%%
In both profiles, isothermal and adiabatic, the EoSs for the description of the crust $(n_{b}< 0.08~{\rm fm^{-3}})$ are provided by the tabulated EoSs with finite temperature from StellarCollapse~\footnote{\url{https://www.stellarcollapse.org}}. In isothermal profile, for the crust region we employed the EoSs of Lattimer and Swesty~\cite{Lattimer-91}. In all cases, the proton fraction of the crust remains a constant value, $Y_{p}=[0.1,0.2,0.3]$, leading to a total of nine EoSs. In adiabatic profile, for the crust region we employed the EoSs of Lattimer and Swesty~\cite{Lattimer-91} and Shen \textit{et al.}~\cite{Shen-2011}. In this case, the proton fraction matches the one of the core EoS, leading to a total of six EoSs.

In addition, for completeness, we employed also the EoSs of Lattimer and Swesty~\cite{Lattimer-91}, Shen \textit{et al.}~\cite{Shen-2011}, Banik \textit{et al.}~\cite{Banik-2014}, and Steiner \textit{et al.}~\cite{Steiner-2013} at full range. More specifically, the Lattimer and Swesty~\cite{Lattimer-91} EoS is applied in $T=[0.01,0.1,1,5,10]~{\rm MeV}$, while the remaining ones, are applied in $T=[0.01,0.1,1]~{\rm MeV}$. For the total of 14 EoSs the proton fraction is constant, $Y_{p}=0.1$.

%%%%%%%%%%%%%%%%%%%%%%%%%%%%%%%%%%%%%%%%%%%%%%%%%%%%%%%%%%%%%%%%
\section{Tidal deformability} \label{sec:tidal}
%%%%%%%%%%%%%%%%%%%%%%%%%%%%%%%%%%%%%%%%%%%%%%%%%%%%%%%%%%%%%%%%%%%%%%%%%%%%%%%%%%%%%%%%%%%%%%%%%%%%%%%%%%
The emitted gravitational waves from the late phase of the inspiral, before the merger, are a very important source for the detectors~\cite{Postnikov-2010,Flanagan-08,Hinderer-08}, leading to the measurement of various properties. During this phase, the tidal effects can be detected~\cite{Flanagan-08}.

The tidal Love number $k_2$, which depends on the EoS, describes the response of a neutron star to the presence of the tidal field $E_{ij}$~\cite{Flanagan-08}. This relation is given below
\begin{equation}
Q_{ij}=-\frac{2}{3}k_2\frac{R^5}{G}E_{ij}\equiv- \lambda E_{ij},
\label{Love-1}
\end{equation}
where $R$ is the neutron star radius and $\lambda=2R^5k_2/3G$ is the tidal deformability. The tidal Love number $k_2$ is given by \cite{Flanagan-08,Hinderer-08}
\begin{eqnarray}
k_2&=&\frac{8\beta^5}{5}\left(1-2\beta\right)^2\left[2-y_R+(y_R-1)2\beta \right]\nonumber\\
& \times&
\left[\frac{}{} 2\beta \left(6  -3y_R+3\beta (5y_R-8)\right) \right. \nonumber \\
&+& 4\beta^3 \left.  \left(13-11y_R+\beta(3y_R-2)+2\beta^2(1+y_R)\right)\frac{}{} \right.\nonumber \\
&+& \left. 3\left(1-2\beta \right)^2\left[2-y_R+2\beta(y_R-1)\right] {\rm ln}\left(1-2\beta\right)\right]^{-1}
\label{k2-def}
\end{eqnarray}
where $\beta=GM/Rc^2$ is the compactness of a neutron star. The quantity $y_R$ is determined by solving the following differential equation
\begin{equation}
r\frac{dy(r)}{dr}+y^2(r)+y(r)F(r)+r^2Q(r)=0, 
\label{D-y-1}
\end{equation}
with the initial condition $ y(0)=2$~\cite{Hinderer-10}. $F(r)$ and $Q(r)$ are functionals of the energy density ${\cal E}(r)$, pressure $P(r)$, and mass $M(r)$ defined as~\cite{Postnikov-2010}
\begin{equation}
F(r)=\left[ 1- \frac{4\pi r^2 G}{c^4}\left({\cal E} (r)-P(r) \right)\right]\left(1-\frac{2M(r)G}{rc^2}  \right)^{-1},
\label{Fr-1}
\end{equation}
and
\begin{eqnarray}
r^2Q(r)&=&\frac{4\pi r^2 G}{c^4} \left[5{\cal E} (r)+9P(r)+\frac{{\cal E} (r)+P(r)}{\partial P(r)/\partial{\cal E} (r)}\right]
\nonumber\\
&\times&
\left(1-\frac{2M(r)G}{rc^2}  \right)^{-1}- 6\left(1-\frac{2M(r)G}{rc^2}  \right)^{-1} \nonumber \\
&-&\frac{4M^2(r)G^2}{r^2c^4}\left(1+\frac{4\pi r^3 P(r)}{M(r)c^2}   \right)^2\left(1-\frac{2M(r)G}{rc^2}  \right)^{-2}.
\label{Qr-1}
\end{eqnarray}
Eq.~(\ref{D-y-1}) must be solved numerically and self consistently with the Tolman - Oppenheimer - Volkoff (TOV) equations using the boundary conditions $y(0)=2$, $P(0)=P_c$ ($P_{c}$ denotes the central pressure), and $M(0)=0$~\cite{Postnikov-2010,Hinderer-08}. From the numerical solution of TOV equations, the mass $M$ and radius $R$ of the neutron star can be extracted, while the corresponding solution of the differential Eq.~(\ref{D-y-1}) provides the value of $y_R=y(R)$. This parameter along with the quantity $\beta$ are the  basic ingredients  of the tidal Love number $k_2$.

The chirp mass {\it $\mathcal{M}_c$} of a binary neutron stars system is a well measured quantity by the detectors~\cite{Abbott-gw170817}. Its exact form is given below
\begin{equation}
\mathcal{M}_c=\frac{(m_1m_2)^{3/5}}{(m_1+m_2)^{1/5}}=m_1\frac{q^{3/5}}{(1+q)^{1/5}},
\label{chirpmass}
\end{equation}
where $m_1$ is the mass of the heavier component star and $m_2$ is the lighter's one. Therefore, the binary mass ratio $q=m_2/m_1$ lies within the range $0\leq q\leq1$.

Also, another quantity that is well measured is the effective tidal deformability $\tilde{\Lambda}$ which is given by~\cite{Abbott-gw170817}
\begin{equation}
\tilde{\Lambda}=\frac{16}{13}\frac{(12q+1)\Lambda_1+(12+q)q^4\Lambda_2}{(1+q)^5},
\label{L-tild-1}
\end{equation}
%%%%%%%%%%%%%%%%%%
where $\Lambda_i$ is the dimensionless deformability~\cite{Abbott-gw170817}
\begin{equation}
\Lambda_i=\frac{2}{3}k_2\left(\frac{R_i c^2}{M_i G}  \right)^5\equiv\frac{2}{3}k_2 \beta_i^{-5}  , \quad i=1,2.
\label{Lamb-1}
\end{equation}
The effective tidal deformability $\tilde{\Lambda}$ plays important role on the neutron star merger process and is one of the main quantities that can be inferred by the detection of the corresponding gravitation waves.
%%%%%%%%%%%%%%%%%%%%%%%%%%%%%%%%%%%
\section{Results and Discussion} \label{sec3:Results}
%%%%%%%%%%%%%%%%%%%%%%%%%%%%%%%%%
In order to study the thermal  effects on the tidal deformability, we employ a class of hot EoSs derived in the past. All of them have been applied successfully in the study of bulk properties of  hot neutron stars, proto-neutron stars, and supernovae. It is worth pointing out that thermal effects are more pronounced in the case of low neutron star mass, which corresponds to the region of the crust, while the effects on the core matter, at least up to 1 MeV, are negligible. We expect that this particular effect, due to quantum behavior of matter, may be reflected not only on the mass-radius diagram of neutron stars for a specific hot EoS, but also on some specific properties that are sensitive on the structure of the neutron star, such as the tidal deformability.

The EoSs employed for the effect of temperature on tidal deformability are based on the work of Ref.~\cite{Koliogiannis-2021}. In particular, we follow the procedure of Ref.~\cite{Koliogiannis-2021} for the construction of the core of the neutron star using the MDI+APR1 EoS, both for isothermal and adiabatic profiles. It has to be noted that the EoSs in both profiles contain only protons, neutrons, and electrons. For the crust of isothermal neutron stars we employed the EoSs of Lattimer and Swesty~\cite{Lattimer-91}, while for the adiabatic ones, we employed the EoSs of Lattimer and Swesty~\cite{Lattimer-91} and Shen \textit{et al.}~\cite{Shen-2011}. More specifically, in isothermal EoSs, the tabulated finite temperature EoSs contain entries up to $n_b = 10^{-13}~{\rm fm^{-3}}$, which in this case, we consider to be the surface of the star ($n_{\rm surf}^{iso} = 10^{-13}~{\rm fm^{-3}}$). However, in the adiabatic EoSs, since the desirable entropy per baryon can be found at various densities, we chose a common value, $n_b = 10^{-15}~{\rm fm^{-3}}$, and extrapolate the EoSs for the crust until this value, using linear dependencies, $log(n_{b})-log(\mathcal{E})$ and $log(n_{b})-log(P)$. In this way, the surface of the adiabatic EoSs is located at $n_{\rm  surf}^{ise} = 10^{-15}~{\rm fm^{-3}}$.

\begin{figure}
\centering
	\includegraphics[width=80mm, height=70mm]{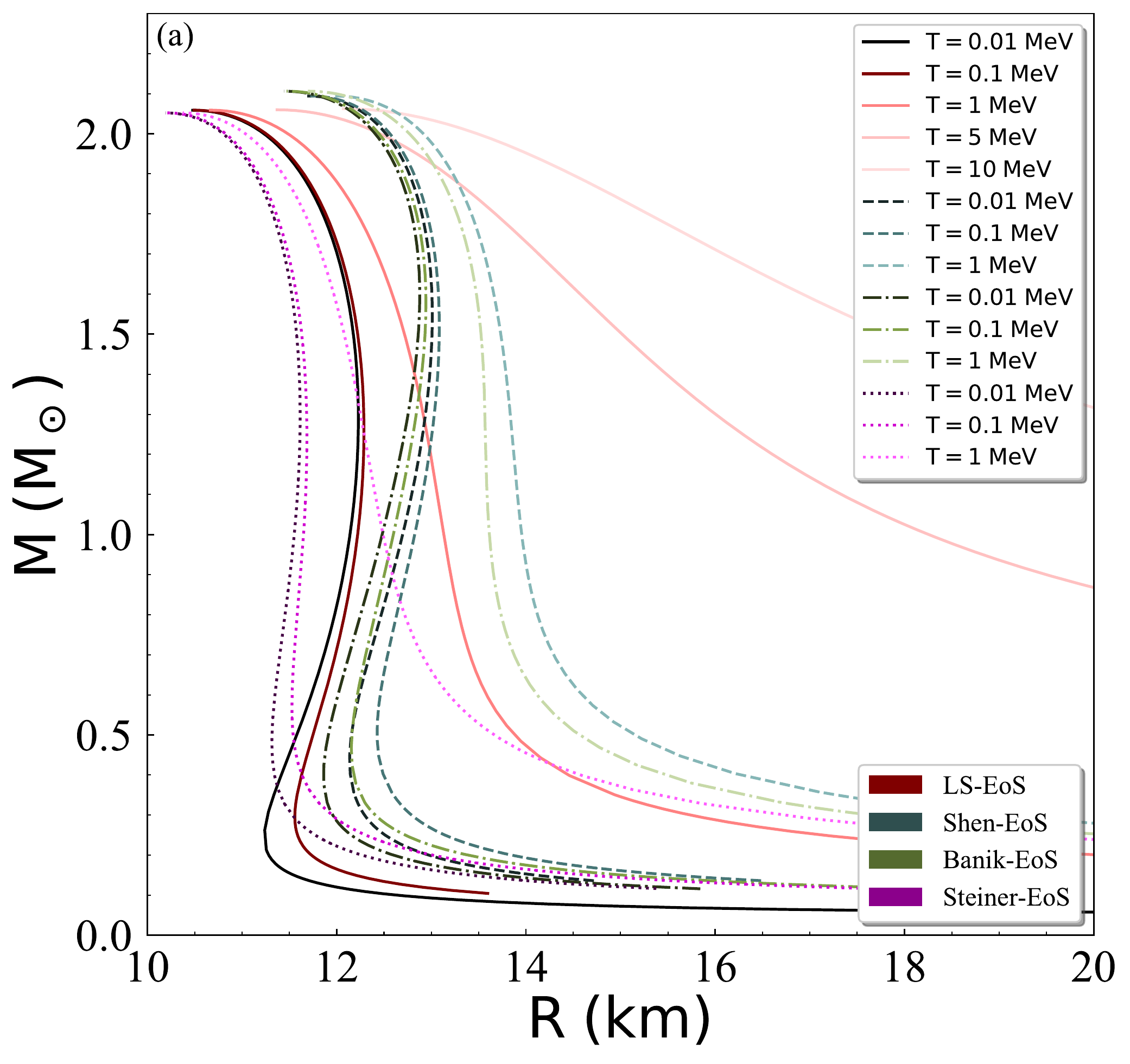}
	\includegraphics[width=80mm, height=70mm]{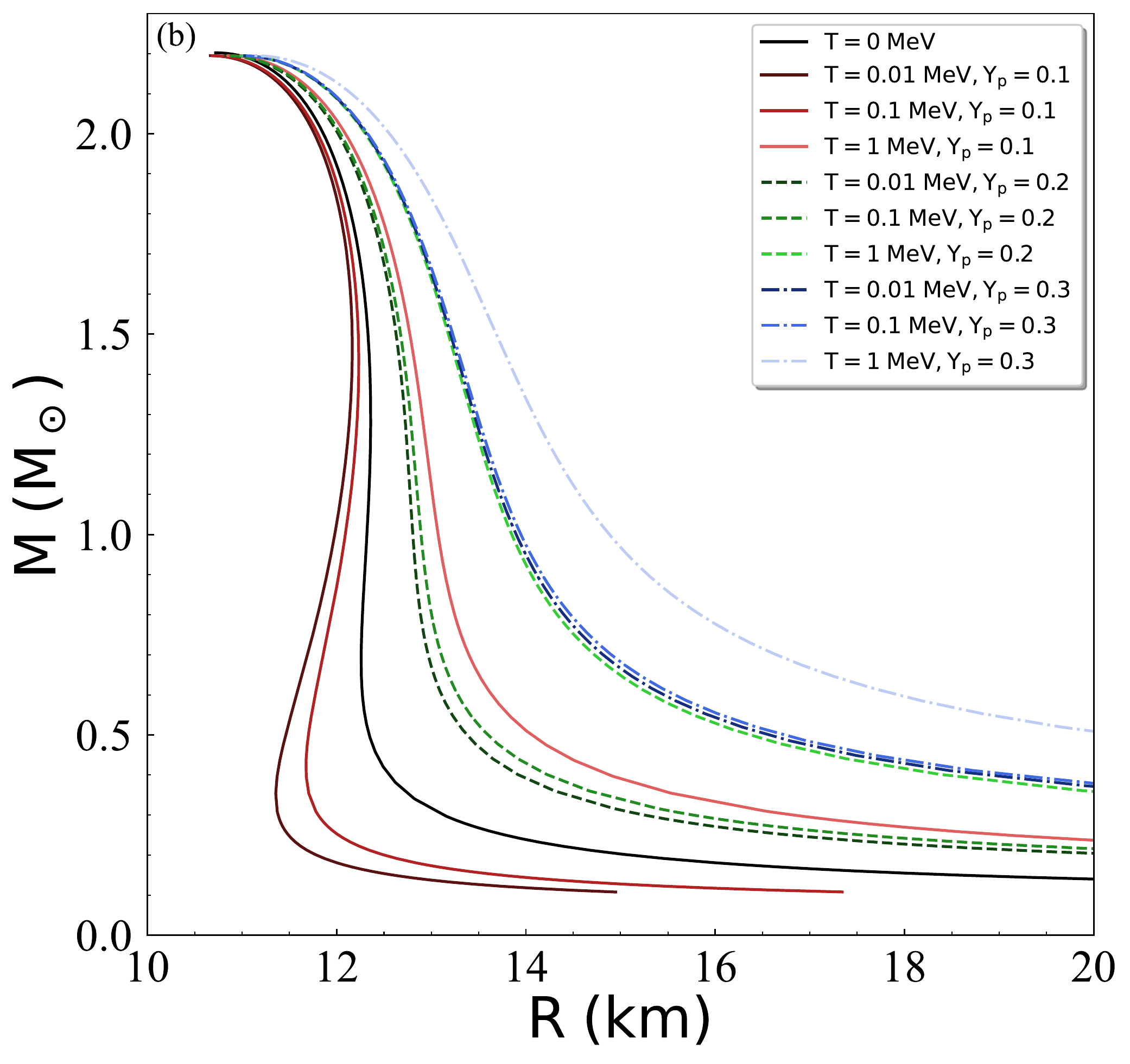}
	\caption{Mass-radius dependence for (a) the four different nuclear models for various values of temperature and (b) various values of temperature and proton fraction for the MDI+APR1 EoS. In panel (a), the red family of curves corresponds to the Lattimer and Swesty EoSs~\cite{Lattimer-91}, the blue family corresponds to the Shen {\it et al.} EoSs~\cite{Shen-2011}, the green family corresponds to the Banik \textit{et al.}  EoSs~\cite{Banik-2014}, and the purple family corresponds to the Steiner \textit{et al.} EoSs~\cite{Steiner-2013}.}
	\label{MRall}
\end{figure}

In particular, we first used the hot EoSs of Lattimer and Swesty~\cite{Lattimer-91}, Shen \textit{et al.}~\cite{Shen-2011}, Banik \textit{et al.}~\cite{Banik-2014}, and Steiner \textit{et al.}~\cite{Steiner-2013}. Moreover, we also employed the MDI+APR1 EoS~\cite{Koliogiannis-2021}. The corresponding model in the latter case has been applied successfully in similar studies and it is suitable for the description of both cold and hot neutron star matter~\cite{Koliogiannis-2021,Koliogiannis-2020}.

In Fig.~\ref{MRall} we display the mass-radius dependence for (a) the four different nuclear models and various values of temperature (the different line styles and family colors correspond to the different models, while the higher temperature corresponds to the lighter color in each case; the same holds for the remaining isothermal figures) and (b) the MDI+APR1 EoS for various values of temperature and proton fraction. In both cases the effect of the temperature is to increase the values of the radius (for a fixed value of mass). This effect is more pronounced at very high values of temperature. Moreover, the increase of the proton fraction on the warm crust leads to the increase of the size of the neutron star, enhancing the effect of temperature.

\begin{table}
\caption{Values of radius and tidal parameters related to the temperature, for a $1.4\;M_\odot$ neutron star and for the Lattimer and Swesty EoSs~\cite{Lattimer-91}.}
\begin{center}
\vspace{0.2cm}
    \begin{tabular}{lccc}
    \hline
    $T$ (MeV) & $k_2$ & $R$ (km) &  $\lambda$ (10$^{36}$gr cm$^2$ s$^2$) \\
    \hline
        0.01    &   0.1005       &    12.21  &     2.73          \\
        0.1   &   0.0984      &   12.26    &  2.73            \\
        1   &   0.0788      &   12.82    &  2.73             \\
        5   &   0.0315      &   15.48    &  2.79             \\
        10   &   0.0127      &   18.92    &  3.02             \\
    \hline
    \end{tabular}
\end{center}
\label{temp-sigma}
\end{table}

\begin{figure*}
\vspace{-0.4cm}
\centering
	\includegraphics[width=155mm]{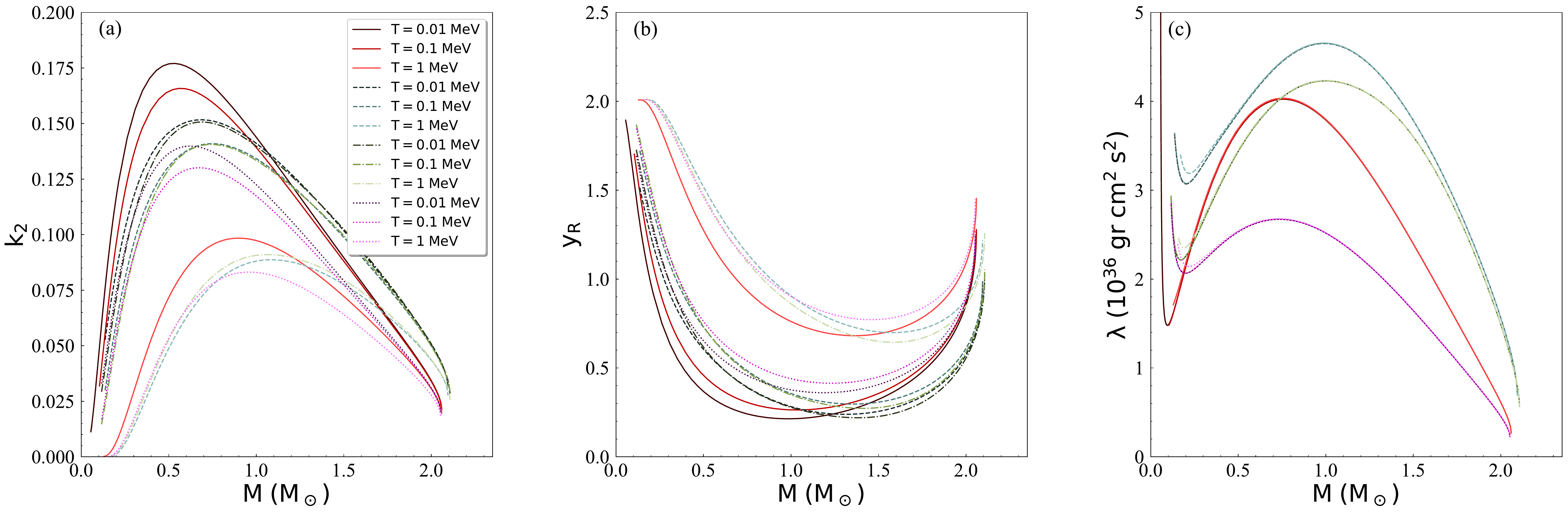}
	~
	\includegraphics[width=155mm]{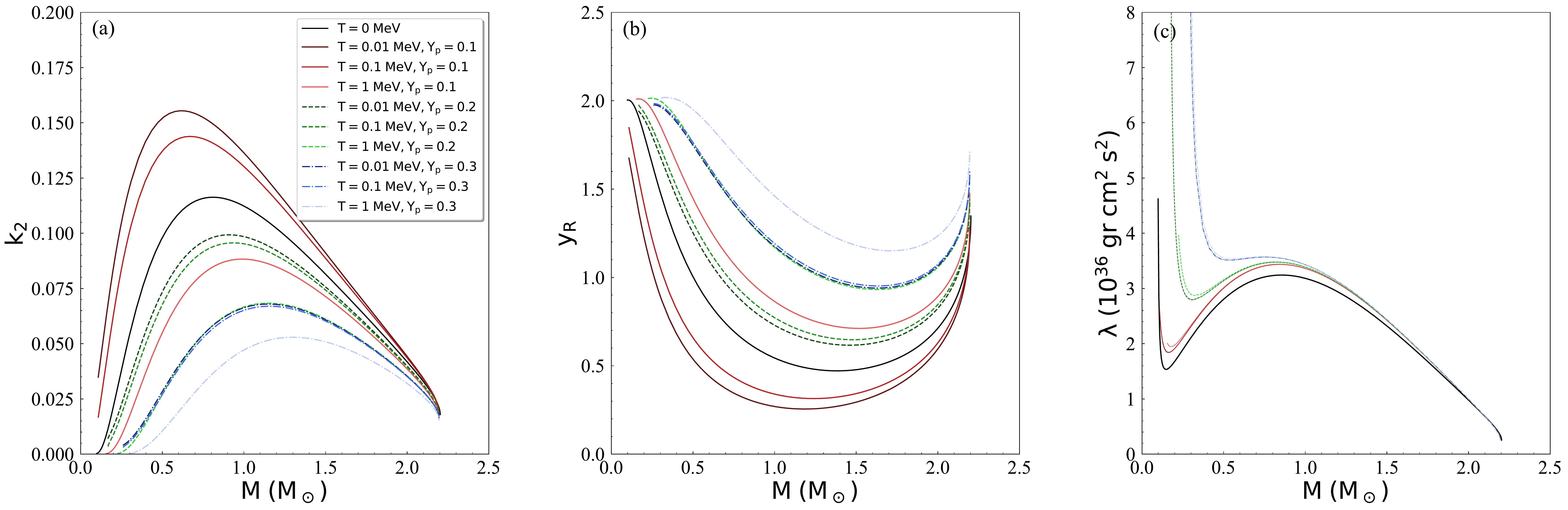}
	\caption{(Top panel) Thermal effects on the Love number $k_2$, parameter $y_R$, and individual tidal deformability $\lambda$ for the four different nuclear models and for various values of temperature. The curves of panels (b) and (c) correspond to the legend of the panel (a). The red family of curves corresponds to the Lattimer and Swesty EoSs~\cite{Lattimer-91}, the blue family corresponds to the Shen {\it et al.} EoSs~\cite{Shen-2011}, the green family corresponds to the Banik \textit{et al.}  EoSs~\cite{Banik-2014}, and the purple family corresponds to the Steiner \textit{et al.} EoSs~\cite{Steiner-2013}. (Bottom panel) Thermal and proton fraction effects on the Love number $k_2$, parameter $y_R$, and individual tidal deformability $\lambda$ for various values of temperature and proton fraction for the MDI+APR1 EoS~\cite{Koliogiannis-2021}. The curves of panels (b) and (c) correspond to the legend of the panel (a).}
	\label{tidalall}
\end{figure*}

Moreover, in Fig.~\ref{tidalall} we display the effects of temperature and proton fraction on the basic ingredients, $k_2$  and $y_R$, of the tidal deformability as well as on $\lambda$ for both the four nuclear models (top panel) and the MDI+APR1 EoS (bottom panel). From Fig.~\ref{tidalall} (top panel), it can be easily seen that, for any  specific EoS, the increase of temperature leads to a decrease in $k_2$, and a corresponding increase on the parameter $y_R$, which is consistent with the structure of the star. In particular, the parameter $y_R$ depends mainly on the neutron star structure, and the increment of the temperature shifts the curves to higher values (reverse behavior compared to $k_2$). However, in Fig.~\ref{tidalall}(c) on the top panel is displayed the surprising result that, although the radius and the Love number $k_2$ are sensitive to temperature, the corresponding tidal deformability $\lambda$ is not, both for low and high values of mass (especially close to $1.4 M_{\odot}$ which is related with the observations). Clearly there is no theoretical evidence for this behavior. In order to further clarify this point and may shed light on the insensitivity of $\lambda$ to the temperature, we consider that in general, $\lambda\sim k_2 R^5 $ holds. The expansion of Love number $k_2$ up to the second order on the compactness parameter $\beta$ reads as~\cite{Piekarewicz-019}
\begin{eqnarray}
k_2&=& -\frac{1}{2}\frac{(y_R-2)}{(y_R+3)}+\frac{5}{2}\frac{(y_R^2+2y_R-6)}{(y_R+3)^2}\beta \nonumber\\
&-& \frac{5}{14}\frac{(11y_R^3+66y_R^2+52y_R-204)}{(y_R+3)^3}\beta^2+{\cal O}(\beta^3).
\end{eqnarray}
Actually, the expansion is accurate considering the Love numbers of neutron stars, even keeping only the first terms. Obviously, there is a   complicated dependence of $k_2$ on $R$ (via $\beta$) and $y_R$. This  makes the dependence of the product $k_2 R^5$ on the temperature quite complex and has no obvious explanation for its independence from the temperature (at least for low temperatures). However, our numerical calculations confirm the insensitive behavior of $\lambda$ on the temperature, at least for low temperatures (see also Table~\ref{temp-sigma}).

Table~\ref{temp-sigma} shows the specific values for the radius, tidal Love number $k_2$ and tidal deformability $\lambda$ for a $1.4\;M_\odot$ neutron star, related to the temperature, considering the Lattimer and Swesty~\cite{Lattimer-91} EoSs (similar results emerge for all the EoSs under consideration). Despite the decrement of $k_2$ and the increment of $R$ as the temperature increases, $\lambda$ remains almost constant in the region $T\in[0.01,1]\;\rm{MeV}$.

\begin{figure*}
\centering
\includegraphics[width=170mm]{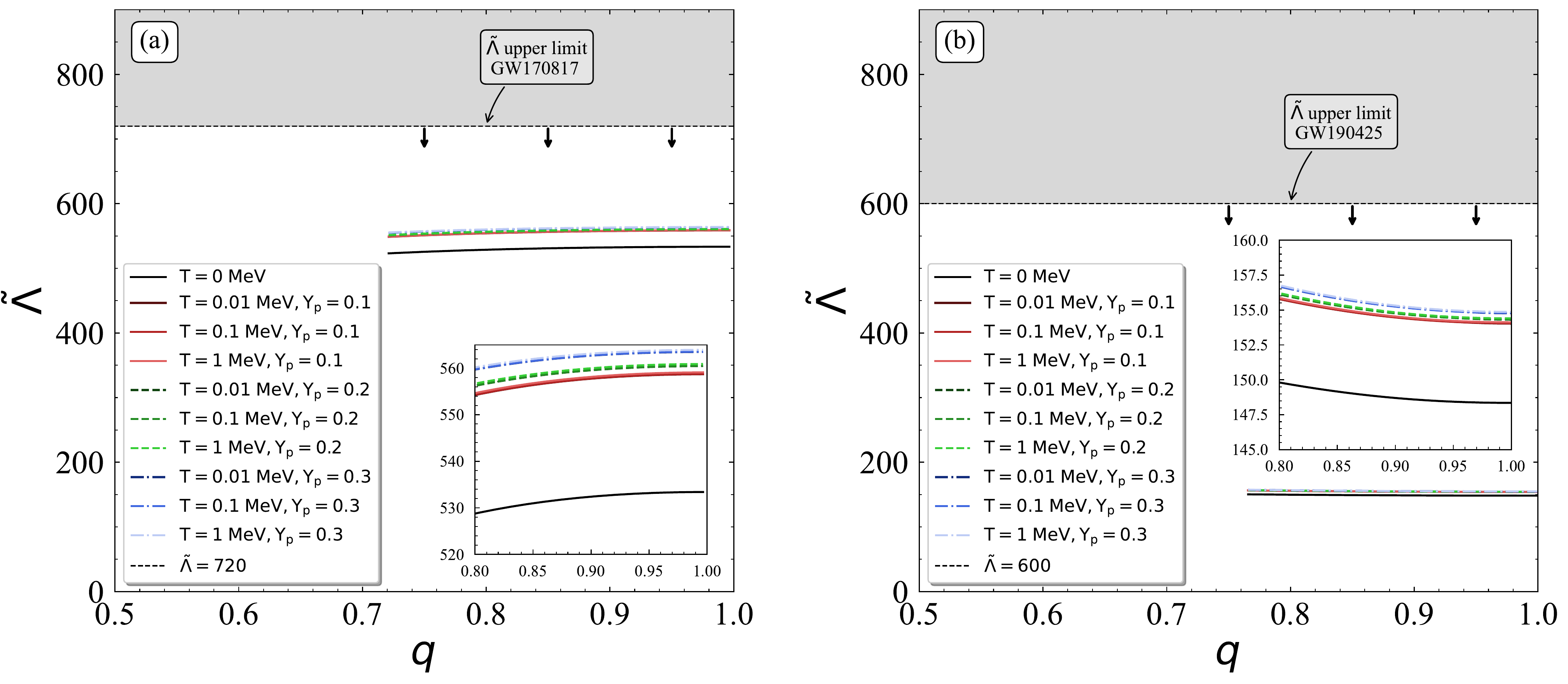}
\caption{Thermal effects on the effective tidal deformability as a function of the mass asymmetry $q=m_2/m_1$ for various values of temperature and proton fraction for the MDI+APR1 EoS and for the (a) GW170817 and (b) GW190425 event, respectively~\cite{Abbott-gw170817,Abbott-gw190425}. The black curve corresponds to the cold EoS. The gray shaded regions indicate the excluded area (the arrows show the accepted region), derived by the measured upper limit on $\tilde{\Lambda}$ for each event.}
\label{Lq}
\end{figure*}

Fig.~\ref{tidalall} (bottom panel) displays the effect of the proton fraction on $\lambda$, which is very small and concerns mainly the very low masses $(M\lesssim 1~M_{\odot})$; a region of no interest in the present study. We notice that despite the differentiation of the EoSs in the panels (a) and (b) for the various values of proton fraction and temperature, this behavior disappears in the $\lambda-M$ relation, leading to an identical behavior of EoSs for high values of mass.

Fig.~\ref{Lq} shows the effect of the temperature and proton fraction on the effective tidal deformability $\tilde{\Lambda}$ for the MDI+APR1 EoS, by using the recent observations of binary neutron star mergers~\cite{Abbott-gw170817,Abbott-gw190425}. In our study we adopted the proposed values for the chirp mass $\mathcal{M}_c$ and the component masses $m_1$, $m_2$ of both events. Specifically, the chirp mass for the GW170817 event is $\mathcal{M}_c=1.186\;M_\odot$~\cite{Abbott-gw170817}, and for the GW190425 is $\mathcal{M}_c=1.44\;M_\odot$~\cite{Abbott-gw190425}. The range for the component masses is $m_1\in(1.36,1.60)\;M_\odot$ and $m_2\in(1.16,1.36)\;M_\odot$~\cite{Abbott-gw170817} (GW170817) and $m_1\in(1.654,1.894)\;M_\odot$ and $m_2\in(1.45,1.654)\;M_\odot$ (GW190425)~\cite{Abbott-gw190425} respectively. We underline that for the mass ranges of the second event we proceeded to a modification, so that $q\leq1$. Fig.~\ref{Lq}(a) corresponds to the GW170817 event while Fig.~\ref{Lq}(b) corresponds to the more recent GW190425 event. Similar to the behavior of the individual tidal deformability $\lambda$ on the bottom panel (c) of Fig.~\ref{tidalall}, the EoSs lead to almost identical behavior, except for a small differentiation from the cold case. Additionally, the effect of the temperature is present mainly on the first event with the lower component masses. Therefore, binary neutron star mergers with a low value of chirp mass $\mathcal{M}_c$ could be more suitable for this kind of study.

In order to further examine thermal  effects on tidal deformability, we employ a set of adiabatic EoSs where we consider that the entropy per baryon is fixed. In this case, the gradient of the temperature is regulated in order to ensure constant entropy in the interior of the star. In particular, we employ two cases of adiabatic EoSs (referring to the selected crust EoS) corresponding to the values $S=[0.1,0.2,0.5]$.

In Fig.~\ref{MRadiabatic}, we display the mass-radius dependence for the two EoSs and the three specific values of the entropy. Thermal effects appear to be insignificant for high values of mass, where for the low values, especially in the range we are interested, that is $M=1.4 \ M_{\odot}$, lead to an increase in the radius of the star. Moreover, in Fig.~\ref{tidalS} we display the effects of entropy on $k_2$, $y_R$, and $\lambda$. Obviously, the thermal effects are negligible on both $k_2$ and $y_R$, and consequently on $\lambda$. Actually, the two cases of EoSs lead to similar predictions for each value of the entropy (the results  are  approximately  independent of the EoS) and depend mainly on the values of the entropy as displayed in Fig.~\ref{tidalS}(b). This remark is well displayed in Table~\ref{adiab-tab}, in which the values of the tidal deformability $\lambda$ are almost identical between the two crust considerations and slightly different for the higher value of the entropy, $S=0.5$, compared to the other two values of $S$.

\setcounter{figure}{3}
\begin{figure}
\centering
\includegraphics[width=80mm]{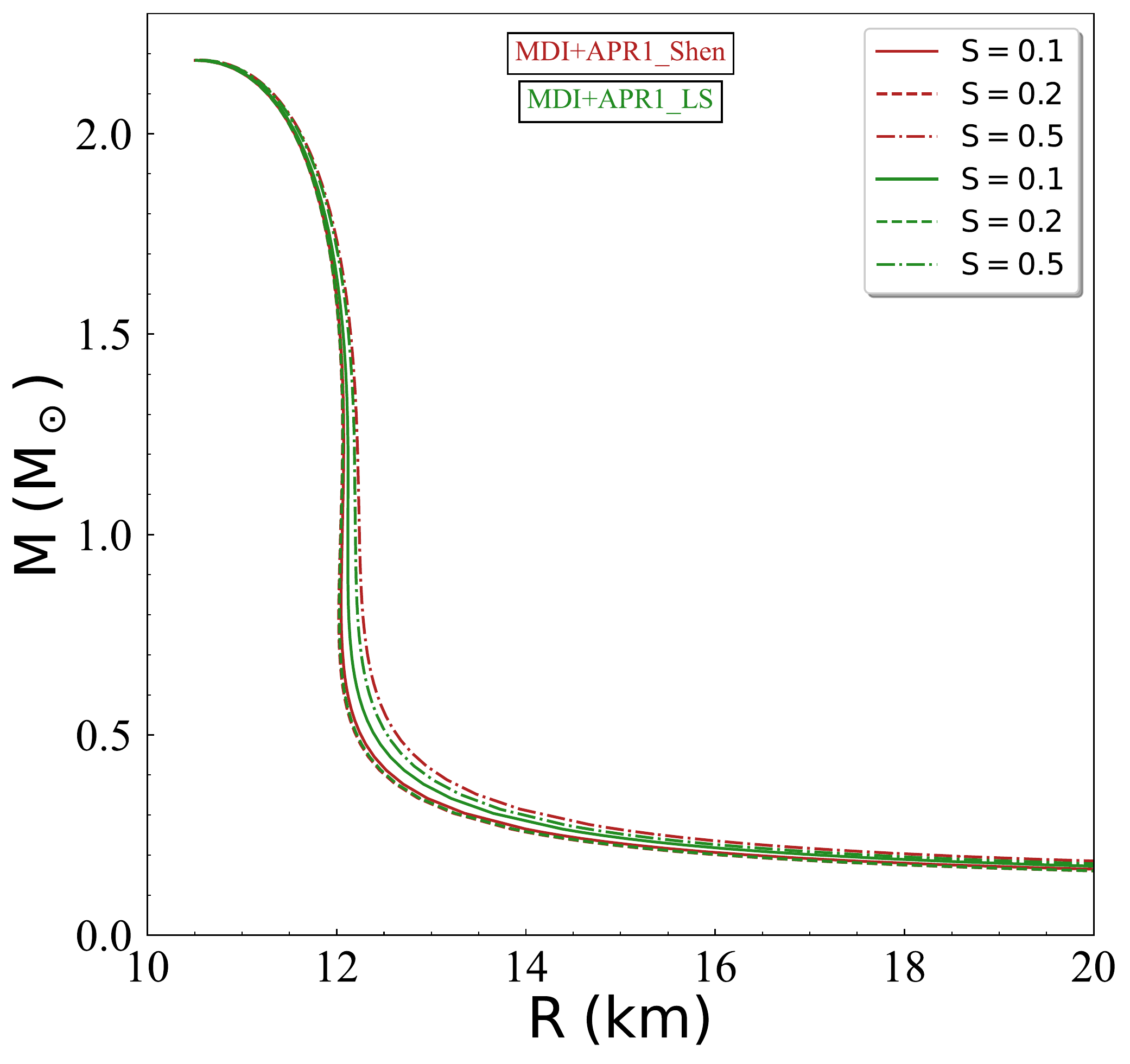}
\caption{Mass-radius dependence for various values of S (in units of ${\rm k}_{B}$) and for both crust considerations. The red (green) curves correspond to the Shen {\it et al.}~\cite{Shen-2011} (Lattimer and Swesty~\cite{Lattimer-91}) crust approach.}
\label{MRadiabatic}
\end{figure}

\setcounter{figure}{4}

\begin{figure*}[!t]
\centering
\includegraphics[width=155mm]{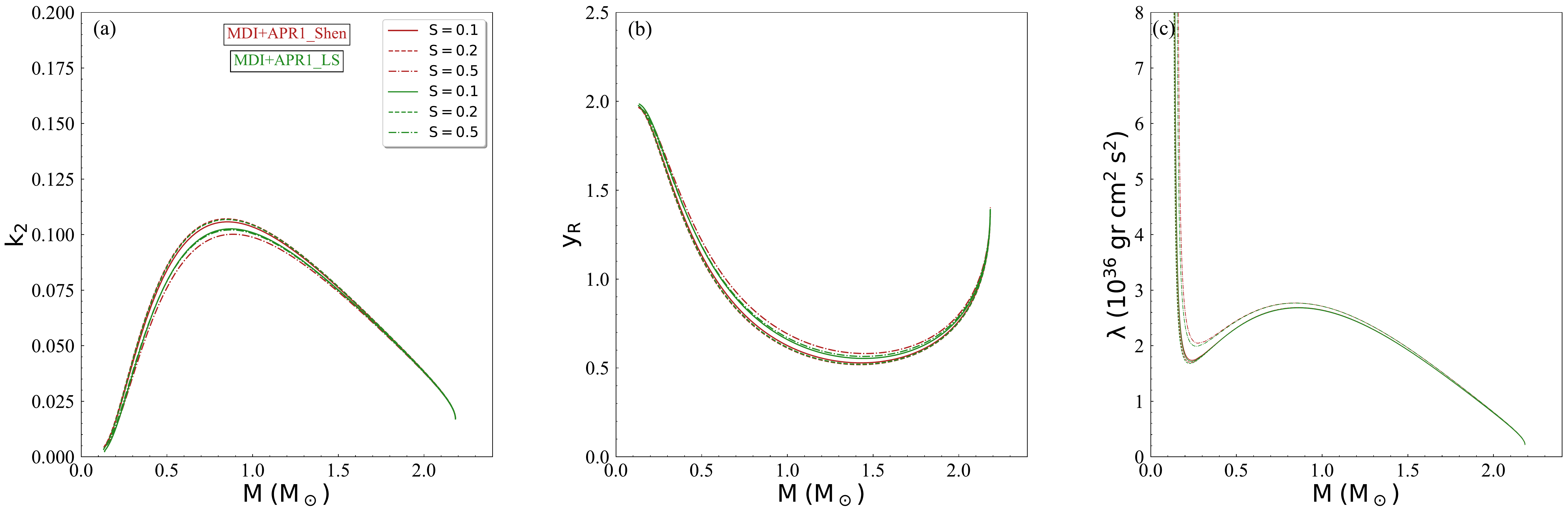}
\caption{Entropy effects on tidal parameters, for both crust considerations. The red (green) curves correspond to the Shen {\it et al.}~\cite{Shen-2011} (Lattimer and Swesty~\cite{Lattimer-91}) crust approach. The curves of panels (b) and (c) correspond to the legend of the panel (a).}
\label{tidalS}
\end{figure*}

\setcounter{figure}{5}
\begin{figure*}[!b]
\centering
\includegraphics[width=170mm]{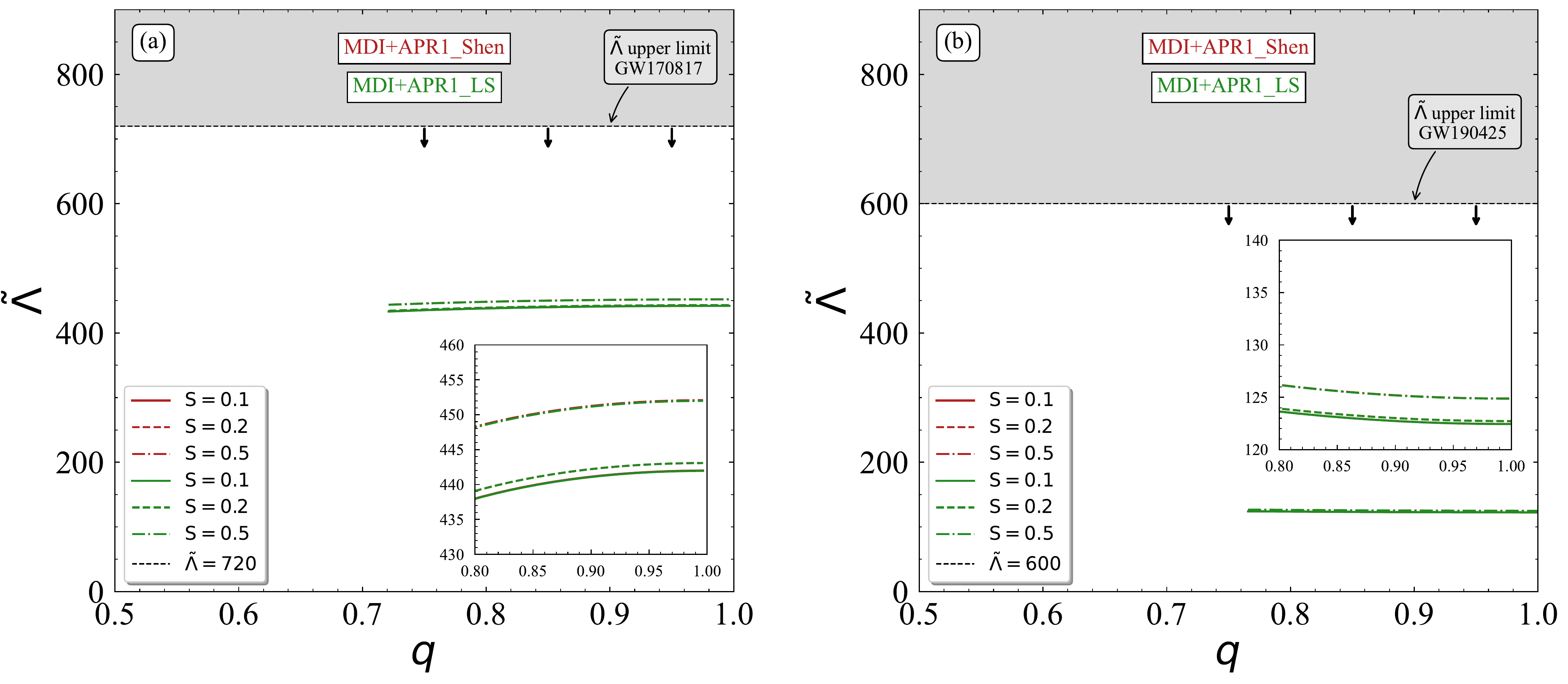}
\caption{Entropy effects on the effective tidal deformability as a function of the mass asymmetry $q=m_2/m_1$, for both crust considerations and (a) GW170817, (b) GW190425 events~\cite{Abbott-gw170817,Abbott-gw190425}. The red (green) curves correspond to the Shen {\it et al.}~\cite{Shen-2011} (Lattimer and Swesty~\cite{Lattimer-91}) crust approach. The gray shaded regions indicate the excluded area (the arrows show the accepted region), derived by the measured upper limit on $\tilde{\Lambda}$ for each event.}
\label{LqS}
\end{figure*}

The above findings are well reflected in Fig.~\ref{LqS}, where we display thermal  effects on the effective tidal deformability $\tilde{\Lambda}$, by using the recent observations of binary neutron star mergers~\cite{Abbott-gw170817,Abbott-gw190425} under the same modification for the range of the component masses with the isothermal case (see Fig.~\ref{Lq} for more details). Firstly, the use of adiabatic EoS leads to lower values of $\tilde{\Lambda}$ in both cases. Moreover, only for high   values of the entropy the effects are important. As a general rule, the use of adiabatic EoSs lead to lower values of $\tilde{\Lambda}$ compared to isothermal ones. In addition, the EoSs are shifted to lower values for the GW190425 event which corresponds to a higher value of chirp mass $\mathcal{M}_c$ (Fig.~\ref{LqS}(b)), compared to the GW170817 event in Fig.~\ref{LqS}(a). Another remark is that binary neutron stars systems with higher value of $\mathcal{M}_c$, such as the GW190425, minimize the differences between the EoSs. This behavior is in accordance to the isothermal case (see Fig.~\ref{Lq}).

Fig.~\ref{R14} demonstrates the relation between the effective tidal deformability $\tilde{\Lambda}$ and the radius $R_{1.4}$ of a $1.4\;M_\odot$ neutron star, by applying the chirp mass $\mathcal{M}_c$ of the GW170817 event. Regarding the isothermal case, the red family of marks corresponds to the Lattimer and Swesty EoSs~\cite{Lattimer-91}, the blue family corresponds to the Shen {\it et al.} EoSs~\cite{Shen-2011}, the green family corresponds to the Banik {\it et al.} EoSs~\cite{Banik-2014}, and the purple family corresponds to the Steiner {\it et al.} EoSs~\cite{Steiner-2013}. As the temperature increases the color of the marks is getting lighter. As a reference we used the range of the component masses of the GW170817 event~\cite{Abbott-gw170817}. The shaded regions correspond to the approximate relations proposed in Ref.~\cite{Zhao}. In particular in Ref.~\cite{Zhao} the authors propose the following dependence between  the radius $R_{1.4}$ and the $\tilde{\Lambda}$
\begin{equation}
\tilde{\Lambda}=a'\left(\frac{R_{1.4}c^2}{G\mathcal{M}_{c}}  \right)^6, \quad a'=0.0035 \pm 0.0007 .
\label{Zhao-form}
\end{equation}

By inverting the Eq.~(\ref{Zhao-form}), the authors proposed the following expression for the $R_{1.4}$
\begin{equation}
    R_{1.4}\simeq(11.5\pm0.3)\frac{\mathcal{M}_{c}}{M_\odot}\left(\frac{\tilde{\Lambda}}{800}\right)^{1/6}\;\;\;\; (\mathrm{km}).
\label{Zhao-expr1}
\end{equation}
In the case of the GW170817 event, the previous relation takes the following form
\begin{equation}
    R_{1.4}\simeq(13.4\pm0.1)\left(\frac{\tilde{\Lambda}}{800}\right)^{1/6}\;\;\;\; (\mathrm{km}).
\label{Zhao-expr2}
\end{equation}

We notice that the adiabatic EoSs lie inside the estimated region of Eq.~(\ref{Zhao-expr2}) (yellow and orange marks in Fig.~\ref{R14}). Also, regarding the isothermal EoSs, we observe that as the temperature increases, the marks are shifted to higher radii. Therefore, if we take into consideration the thermal effects, possible constraints on the radius could lead to constraints on the value of the possibly existing temperature.
%%%%%%%%%%%%%%%%%%%%%%%%%%%%%%%%%%%%%%%%%%%%%%%%%%%%%%%%

\begin{table*}
\caption{Values of radius and tidal parameters related to the entropy, for a $1.4\;M_\odot$ neutron star and for both crust considerations (Shen \textit{et al.}~\cite{Shen-2011}, and Lattimer-Swesty~\cite{Lattimer-91}).}
\begin{center}
\vspace{0.2cm}
\begin{tabular}{ccccccc}
\hline
\multirow{2}{*}{S ($\mathrm{k_B}$)} & \multicolumn{3}{c}{Shen {\it et al.}~\cite{Shen-2011}} & \multicolumn{3}{c}{Lattimer and Swesty~\cite{Lattimer-91}} \\ \cline{2-7}
&   $k_2$       &    $R$ (km)   &  $\lambda$ (10$^{36}$gr cm$^2$ s$^2$)    &   $k_2$       &    $R$ (km)   &  $\lambda$ (10$^{36}$gr cm$^2$ s$^2$)        \\ \hline
0.1 & 0.0830 & 12.06 & 2.114 & 0.0817 & 12.10 & 2.114 \\
0.2 & 0.0835 & 12.05 & 2.119 & 0.0834 & 12.05 & 2.119 \\
0.5 & 0.0806 & 12.18 & 2.162 & 0.0814 & 12.16 & 2.161 \\ \hline
\end{tabular}
\end{center}
\label{adiab-tab}
\end{table*}

%%%%%%%%%%%%%%%%%%%%%%%%%%
\section{Conclusions} 
\label{sec3:Conclus}
%%%%%%%%%%%%%%%%%%%%%%%%%%%
In the majority of the recent and previous studies, the tidal deformability is calculated considering the cold EoS for the description of the interior of the neutron stars before the merger. According to our knowledge, this is the  first effort  dedicated in studying thermal effects on tidal deformability (and consequently on the gravitation waves signal) just before the merger. Although there are some theoretical arguments supporting the idea of cold neutron stars before the merger (temperature less than $0.01$ MeV), other studies claim that it is possible for both stars to warm enough, with their interior reaching temperatures even a few MeV. In any case, it is worth studying and also comparing the two possibilities in order to gain useful insight. Moreover, it is useful to enrich our knowledge on the effect of temperature on tidal deformability, since this quantity is mainly related to the detection and analysis of gravitational waves.

\begin{figure}
\centering
\includegraphics[width=80mm]{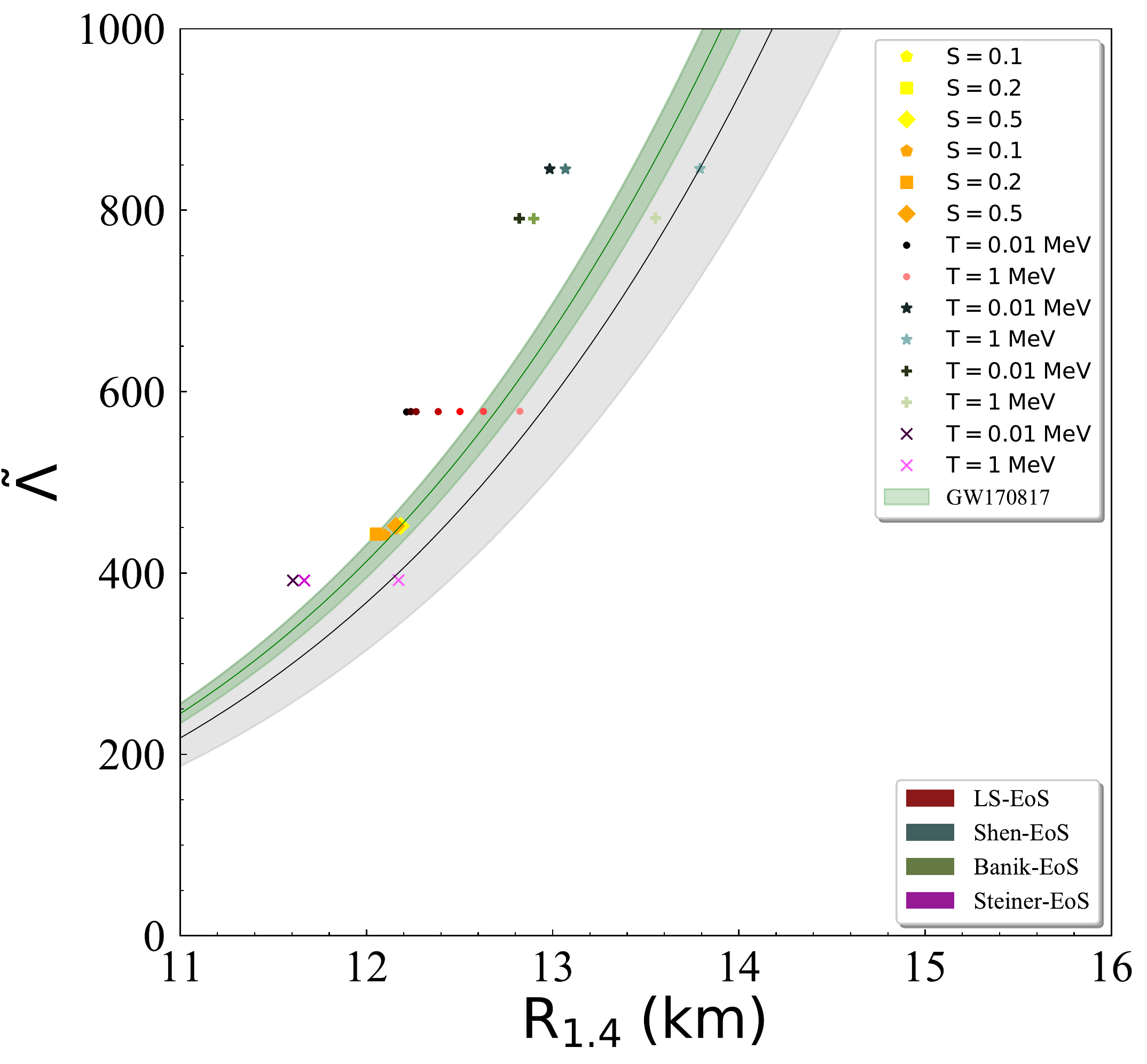}
\caption{$\tilde{\Lambda}$ versus radius $R_{1.4}$ for the four different nuclear models (isothermal case). Especially, the red family of curves corresponds to the Lattimer and Swesty EoSs~\cite{Lattimer-91}, the blue family corresponds to the Shen {\it et al.} EoSs~\cite{Shen-2011}, the green family corresponds to the Banik {\it et al.} EoSs~\cite{Banik-2014}, and the purple family corresponds to the Steiner {\it et al.} EoSs~\cite{Steiner-2013}. The yellow (orange) marks correspond to the Shen \textit{et al.}~\cite{Shen-2011} (Lattimer and Swesty~\cite{Lattimer-91}) consideration for the crust of the adiabatic case. The gray (green) shaded area and green (black) line correspond to the proposed expression of Eq.~(\ref{Zhao-expr1}) [Eq.~(\ref{Zhao-expr2})].}
\label{R14}
\end{figure}

In this research we used various sets of hot EoSs predicted in previous studies, as well as a hot EoS suitable to describe the interior of a neutron star even for very high temperatures. We found the unexpected result, at least in terms of some solid theoretical proof, that although thermal effects for low values of $T$ ($T<1$ MeV) are important for  $k_2$ and $R$, that is not the case for $\lambda \sim k_2 R^5$ (or $\Lambda$). The present study leads to the conclusion that the above rule applies regardless of the applied EoS. If this estimation proves to be correct, then it will have consequences in the way we draw conclusions from the observation of gravitational waves. The accurate measurement of the radius, with some other reliable method can provide information about the temperature of the stars in the phase before the merger (see discussion of Fig.~\ref{R14}).

Concluding, we expect future observations of gravitational waves of neutron stars' merger events may shed light on this problem. Particularly, stringent constraints on $\tilde{\Lambda}-R$ dependence may help to clarify the extent of the thermal effects on the coalescing neutron stars and vice-versa. It is among our immediate intentions to study more systematically the effect of temperature on tidal deformability using a broader and more elaborated  set of hot EoSs. In addition, we are working on a detailed study for the role of the hot crust, whose structure and composition is  more sensitive to the effect of the temperature, due to its lower values of density (compared to the core).

%%%%%%%%%%%%%%%%%%%%%%%%%
%\appendix
%\section*{Appendix}
%%%%%%%%%%%%%%%%%%%%%%%%%%%%

%%%%%%%%%%%%%%%%%%%%%%%%%%%%%
\section*{Acknowledgments}
%%%%%%%%%%%%%%%%%%%%%%%%%%%%%%
We would like to thank Prof. P. Meszaros, Dr. S. Typel and Prof. D. Radice for their useful corresponds and suggestions. The research work was supported by the Hellenic Foundation for Research and Innovation (HFRI) under the 3rd Call for HFRI PhD Fellowships (Fellowship Number: 5657).





\end{document}